\documentclass[prb,amsmath,showpacs]{revtex4}
\usepackage{latexsym}
\usepackage{graphicx}

\begin{document}
\title{Investigation of the oxohalogenide Cu$_{4}$Te$_{5}$O$_{12}$Cl$_{4}$
with weakly coupled Cu(II) tetrahedra}

\author {
 Rie Takagi$^{a}$,
 Mats Johnsson$^{a}$,
 Vladimir Gnezdilov$^{b}$,
 Reinhard K. Kremer$^{c}$,
 Wolfram Brenig$^{d}$,
 and Peter Lemmens$^{c,e}$}

\affiliation{
 $^{a}$ Department of Inorganic Chemistry, Stockholm University, S-106 91
Stockholm, Sweden\\
 $^{b}$ B. I. Verkin Institute for Low Temperature Physics NASU, 61164 Kharkov,
Ukraine\\
 $^{c}$ Max Planck Institute for Solid State Research, Heisenbergstrasse 1,
D-70569 Stuttgart, Germany\\
$^{d}$ Institute for Theoretical Physics, TU Braunschweig,
D-38106 Braunschweig, Germany\\
 $^{e}$ Institute for Physics of Condensed Matter, TU Braunschweig,
D-38106 Braunschweig, Germany}\

\date{\today}

\begin{abstract} The crystal structure of the copper(II) tellurium(IV) oxochloride
Cu$_{4}$Te$_{5}$O$_{12}$Cl$_{4}$ (Cu-45124) is composed of weakly coupled tetrahedral Cu
clusters and shows crystallographic similarities with the intensively investigated
compound Cu$_{2}$Te$_{2}$O$_{5}$X$_{2}$, with X~=~Cl, Br (Cu-2252). It differs from the
latter by a larger separation of the tetrahedra within the crystallographic \emph{ab}
plane, that allows a more direct assignment of important inter-tetrahedra exchange paths
and the existence of an inversion center. Magnetic susceptibility and specific heat
evidence antiferromagnetic, frustrated correlations of the Cu spin moments and long
range ordering with $T_{c}$=13.6 K. The entropy related to the transition is reduced due
to quantum fluctuations. In Raman scattering a well structured low energy magnetic
excitation is observed at energies of $\approx$50K (35cm$^{-1})$. This energy scale is
reduced as compared to Cu-2252.
\end{abstract}


\maketitle


\section{Introduction}
Quantum spin systems are interesting due to strong fluctuation effects, unconventional
ground states and a pronounced electronic and magnetic softness. The latter leads to
large responses to external fields and changes of composition \cite{dagotto02}.
Competing exchange interactions in geometrically frustrated, e.g. triangular or
tetrahedral lattices, enhance this trend, suppress long range ordering and shift
relevant excitations to lower energies. An important aspect of such systems is the
coupling of the frustrated entities, e.g. triangles or tetrahedra, to larger units, as
planes or framework structures. Quantum criticality and related phenomena can be tuned
by favorably modifying the coupling between such units
\cite{anderson63,mila00,greedan01,lhuillier05}. Strongly coupled, corner sharing
tetrahedra exist e.g. in the pyrochlores. These compounds show spin ice states or
unconventional transitions into spin liquid phases
\cite{ramirez99,bramwell01,lee02,lee04}. The implications of inter-tetrahedra couplings
of different strength, dimensionality or topology are intensively studied theoretically,
for example in Ref. \onlinecite{brenig01,brenig03,brenig04,kotov05}.

The compound Cu$_{2}$Te$_{2}$O$_{5}$X$_{2}$, X=Br and Cl, (Cu-2252) which contains
weakly coupled Cu$^{2+}_4$ tetrahedra is a model system with this respect as the
coupling of the Cu tetrahedra can be tuned continuously by varying the stoichiometry
\cite{johnsson00,lemmens01,lemmens02b}. Cu-2252 shows incommensurate long-range ordering
with strongly reduced ordered moments and transition temperatures for X=Br. The ordered
magnetic structure is most probably a complex helical structure as demonstrated for
X=Cl\cite{zaharko04}. Finally, also unconventional collective modes are observed as
longitudinal magnons for X=Br in Raman scattering
\cite{lemmens01,gros03,lemmens02-proceeding} and a dichotomy of temperature dependent
and invariant magnons in neutron scattering \cite{zaharko04,zaharko05}. The presence of
longitudinal magnons has been taken as evidence for the system being close to a quantum
critical point \cite{lemmens02b,gros03}.

Hydrostatic pressure enhances the intra-tetrahedra coupling in both compounds and
reduces the inter-tetrahedra coupling for X=Br, respectively, shifting the system closer
to the quantum critical point \cite{kreitlow05,crowe06}. There even exist evidence for a
complete suppression of long range order for X=Br\cite{kreitlow05}. The large response
of Cu-2252 on changes of composition and hydrostatic pressure \cite{wang05} is based on
a complex network of exchange paths that are dominated by halogen-mediated exchange in
the \emph{ab} plane \cite{valenti03b}.

The aim of our present study is to extend the number of related systems with unusual
magnetic properties. So far, in the phase diagram
CuO~:~CuX$_{2}$~(X~=~Cl,~Br)~:~TeO$_{2}$ only two compounds have been identified; the
previously discussed Cu-2252 \cite{johnsson00} and a mixed AF/FM dimer chain system
Cu$_{3}$(TeO$_{3})_{2}$Br$_{2}$ \cite{becker05}. The crystal structure of the present
system Cu$_{4}$Te$_{5}$O$_{12}$Cl$_{4}$ (Cu-45124) has many similarities with the
previously know Cu-2252. Nevertheless, their physical properties are different enough to
enable to a better understanding of the properties and phase diagram of weakly coupled
tetrahedra systems.

In the following we will describe aspects of the sample preparation of Cu-45124, compare
its structural and electronic properties with those of  other Te(IV) electron lone pair
systems and investigate its thermodynamic and spectroscopic properties. The relevance of
Dzyaloshinsky-Moriya (DM) and spin-phonon interaction will be discussed.

\section{Experimental}
Single crystals of the compound Cu-45124 were synthesized in sealed
evacuated silica tubes. CuCl$_{2}$ (Avocado Research Chemicals,
+98{\%}), CuO (Avocado Research Chemicals, +99{\%}), TeO$_{2}$
(ABCR, +99{\%}) were used as starting materials. CuCl$_{2}$ : CuO :
TeO$_{2}$ were mixed in the stoichiometric molar ratio 2~:~2~:~5 and
sealed into evacuated silica tubes (length $\sim $ 6 cm).  The tubes
were heated at 500 \r{ }C for 72~h in a box furnace. The product
consists of small green bulky non-hygroscopic single crystals.
Attempts to synthesize a Br-analogue failed so far.  Our experiments
gave the previously known compound Cu$_{2}$Te$_{2}$O$_{5}$Br$_{2}$
in addition to unreacted TeO$_{2}$.

Single-crystal X-ray diffraction data were collected on a STOE IPDS
image-plate rotating anode diffractometer using
graphite-monochromatized Mo K$\alpha $ radiation, $\lambda $ =
0.71073 {\AA}. The intensities of the reflections were integrated
with the STOE software and absorption corrections were carried out
numerically, after crystal shape optimization \cite{x-shape,x-red}.
The structure was solved by direct methods \cite{shelxs} and refined
by full matrix least squares on F$^{2}$ (Ref~\onlinecite{shelxl}).
All atoms were refined with anisotropic displacement parameters. The
 crystallographic data  are reported in
Table~\ref{cryst-data} \cite{supplement}. The chemical composition
was checked in a scanning electron microscope (SEM, JEOL 820)
equipped with an energy-dispersive spectrometer (EDS, LINK AN10000).

Monophasic powder of Cu-45124 was checked with X-ray powder diffraction using a
Guiner-H\"{a}gg focusing camera with subtraction geometry (CuK$\alpha _{1}$ radiation,
$\lambda$~=~1.54060 {\AA}) For the determination of the lattice parameters silicon
($a$~=~5.43088(4)~{\AA}) was added as an internal standard. The recorded films were read
in with an automatic film scanner and the data were treated using the programs SCANPI
and PIRUM. Refinement of the tetragonal unit cell by powder X-ray diffraction resulted
in $a$~=~11.3474(6)~{\AA}, $c$ =~6.3439(5)~{\AA}. The magnetic susceptibility and
specific heat data were collected using a SQUID magnetometer (MPMS, Quantum Design) and
a Physical Property Measurement System system with specific heat options (PPMS, Quantum
Design). For Raman scattering experiments individual single crystals with typical
dimensions of approximately 200-300 $\mu $m diameter were used with the 514.5~nm
excitation line of an Ar$^{+}$ Laser and a laser power of P=1~mW in quasi-backscattering
geometry. The scattered spectra were collected by a DILOR-XY triple spectrometer and a
nitrogen cooled CCD detector with a spectral resolution of approximately 1~cm$^{-1}$.
Due to the transparency and irregular shape of the single crystals the exciting Laser
line probes the bulk of the crystal. However, symmetry information of the excitations is
lost.

\section{Crystal structure}
\begin{figure}[htbp]
\centerline{\includegraphics[height=5.0cm]{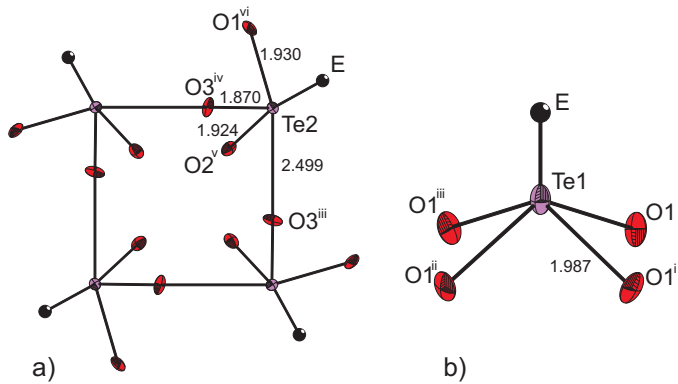}} \caption{(Color online) a) Four
[Te(2)O$_{3+1}$E] polyhedra are connected via corner sharing to form
[Te$_{4}$O$_{12}$E$_{4}$] rings. b) The Te(1) atom is coordinated by four O(1) atoms and
the lone-pair to form a [TeO$_{4}$E] square pyramid. Distances from Te to the different
oxygen are given in {\AA}. } \label{fig1}
\end{figure}

\begin{figure}[htbp]
\centerline{\includegraphics[height=5.5cm]{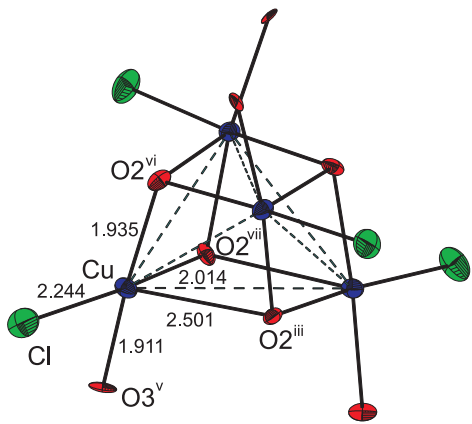}} \caption{(Color online) Groups of
four [CuO$_{4}$Cl] square pyramids are connected via edge sharing to form
[Cu$_{4}$O$_{8}$Cl$_{4}$] units. The Cu$^{2+}$ ions form a distorted tetrahedron marked
with dashed lines. } \label{fig2}
\end{figure}

The present compound Cu-45124 crystallizes in the tetragonal system, space group P4/n
\cite{supplement}. Atomic coordinates and selected angles are listed in the
Tables~\ref{cryst-data}, \ref{coordinates}, \ref{angles} and are shown in the Figures
\ref{fig1} and \ref{fig2}. As the chemical and structural peculiarities of this class of
compounds are based on Te$^{4+}$ and its lone pair electron we will first discuss the Te
coordinations. The interplay of the lone pair electrons on the background of the
oxohalide framework is important as it allows voids in the crystal structure and a large
electronic polarizability. This shows up, e.~g. as a large Raman scattering intensity in
Cu-45124. The intra- and inter-tetrahedra exchange paths and exchange couplings that
realize weakly coupled tetrahedra are dominated by Cu--O and Cu--Cl coordinations and
will be discussed later.

\begin{table}[htbp]
\caption{Crystal data for Cu-45124 at T=298~K \cite{supplement}. }
\begin{center}
\begin{tabular}{|l|l|}
\hline Empirical formula &
Cu$_{4}$Te$_{5}$O$_{12}$Cl$_{4}$ \\
\hline Formula weight & 1225.96 g/mol \\
\hline Crystal system, Space group&
Tetragonal, P 4/n\\
\hline Unit cell dimensions& $a~$=~11.3474(16) {\AA}\\
                           & $c~$=~6.3319(9) {\AA}
\\\hline Volume &
815.3(2) {\AA}$^{3}$ \\
\hline Z& 2 \\
\hline Density (calculated)& 4.994 g/cm$^{3}$ \\\hline
\end{tabular}
\label{cryst-data}
\end{center}
\end{table}

\begin{table}[htbp]
\caption{Atomic coordinates for Cu-45124. All atoms are refined with anisotropic
displacement parameters, however, only the isotropic displacement parameters are shown
in the table. U(eq) is defined as one-third of the trace of the orthogonalized U tensor.
BVS is the bond valence sum calculated using parameters from
Refs.~\onlinecite{brown85,brese91}.}
\begin{center}
\begin{tabular}{|l|l|l|l|l|l|l|}
\hline \textbf{Atom}& \textbf{Wyck.}& \textbf{x}& \textbf{y}& \textbf{z}&
\textbf{U}$_{eq}$\textbf{ [{\AA}}$^{2}$\textbf{]}&
\textbf{BVS }  \\
\hline Te(1)& 2c& 1/4& 1/4& 0.3682(3)& 0.0117(3)~&
3.89 \\
\hline Te(2)& 8g& 0.67623(5)& 0.02031(5)& 0.86708(11)& 0.0075(2)~&
3.87 \\
\hline Cu& 8g& 0.75930(11)& 0.40494(10)& 0.3481(2)& 0.0110(3)&
2.06 \\
\hline Cl& 8g& 0.8874(3)& 0.5554(2)& 0.3341(5)& 0.0225(7)~&
0.52 \\
\hline O(1)& 8g& 0.2867(6)& 0.4044(6)& 0.2355(12)& 0.0138(17)&
2.13 \\
\hline O(2)& 8g& 0.2848(5)& 0.8712(6)& 0.3558(12)& 0.0098(15)&
2.17 \\
\hline O(3)& 8g& 0.2876(6)& 0.5764(6)& 0.9397(11)& 0.0100(15)&
2.11 \\
\hline
\end{tabular}
\label{coordinates}
\end{center}
\end{table}

\begin{table}[htbp]
\caption{Selected bond angles (\r{ }) for Cu-45124.}
\begin{center}
\begin{tabular}{|l|l|l|l|l|}
\hline O(1)$^{}$---Te(1)---O(1)$^{}$$\times$4& 79.69(18)& ~&
O(3)$^{v}$---Cu---O(2)$^{vi}$&
171.1(3) \\
\hline O(1)$^{}$---Te(1)---O(1)$^{}$$\times$2& 130.0(5) & &
O(3)$^{v}$---Cu---O(2)$^{vii}$&
83.9(3) \\
\hline O(3)$^{iv}$---Te(2)---O(2)$^{v}$& 96.1(3) & & O(2)$^{vi}$---Cu---O(2)$^{vii}$&
87.3(3) \\
\hline O(3)$^{iv}$---Te(2)---O(1)$^{vi}$& 90.7(3) & & O(3)$^{v}$---Cu---Cl&
93.3(2) \\
\hline O(2)$^{v}$---Te(2)---O(1)$^{vi}$& 92.3(3) & & O(2)$^{vi}$---Cu---Cl&
95.3(2) \\
\hline O(3)$^{iv}$---Te(2)---O(3)$^{iii}$& 83.6(3) & & O(2)$^{vii}$---Cu---Cl&
173.0(2) \\
\hline O(2)$^{v}$---Te(2)---O(3)$^{iii}$& 71.4(3)& & O(3)$^{v}$---Cu---O(2)$^{iii}$&
104.8(3) \\
\hline O(1)$^{vi}$---Te(2)---O(3)$^{iii}$& {162.0(3)}&& O(2)$^{vi}$---Cu---O(2)$^{iii}$&
74.6(3) \\
\hline                      &         & & O(2)$^{vii}$---Cu---O(2)$^{iii}$& 77.9(3) \\
\hline                      &         & & Cl---Cu---O(2)$^{iii}$& 109.09(17) \\
\hline
\end{tabular}
\label{angles}
\end{center}
\end{table}

%
%

The Te(2) atom has a see-saw [TeO$_{3+1}$] coordination with three Te -- O bond
distances in the range 1.870 -- 1.930 {\AA}, the fourth long Te(2) -- O(3) distance
amounts to 2.499 {\AA}. There is a fifth long Te(2) -- O(1) distance with 2.88 {\AA},
however the oxygen atom O(1) is not considered to belong to the primary coordination
sphere of Te(2) according to the bond valence sum calculations \cite{brown85,brown02}.
The coordination polyhedron with three short Te -- O  and one longer distance is common
for Te$^{4+}$ (Ref.~\onlinecite{zemann71}). The coordinated atoms form a
[Te(2)O$_{3+1}$E] trigonal bipyramid when also the stereochemically active 5s$^{2}$ lone
pair (designated E) located in the equatorial plane is taken into account. Four
[Te(2)O$_{3+1}$E] polyhedra are connected sharing common corners to form
[Te$_{4}$O$_{12}$E$_{4}$] rings, see Figure~\ref{fig1}a. The Te(1) atom is coordinated
by four O(1) atoms to form a [TeO$_{4}$] pyramid, see Figure~\ref{fig1}b. This is a very
uncommon coordination polyhedron for Te$^{4+}$ and to our best knowledge has not been
observed before. When the stereochemically active lone pair is also taken into
consideration the coordinated atoms form a [Te(1)O$_{4}$E] square pyramid. The Te(1) --
O bonding distance is 1.987~{\AA} and the O(1)~--~Te(1)~--~O(1) angles amount to
79.69(18)$^{o}$. This type of coordination has previously been observed for Sb$^{3+}$,
and Bi$^{3+ }$ in, $e.g.$ in the compounds BaSbO$_{2}$Cl
(Ref.~\onlinecite{thuillier80}), DyBi$_{2}$O$_{4}$I (Ref.~\onlinecite{schmidt00}).

The Cu atom is coordinated by four oxygen atoms and one chlorine atom to form a
distorted [CuO$_{4}$Cl] square pyramid. Three Cu -- O bonding distances are in the range
1.911 -- 2.014~{\AA} and one long is present at 2.501 {\AA}. The Cu -- Cl bond distance
is 2.244~{\AA}. These coordinations leads to a Cu bond valence sum of +2.06 (see Table
\ref{coordinates}). Each Cl atom forms only one bond that leads to voids in the
structure along [001] where the Cl atoms and the Te(1) lone pair are located. Therefore
the bond valence sum for the Cl$^{-}$ ion in Cu-45124 is only 0.52 suggesting that it
takes the role of a counter ion instead of being fully integrated in the covalent/ionic
network.

\begin{figure}[htbp]
\centerline{\includegraphics[height=5.5cm]{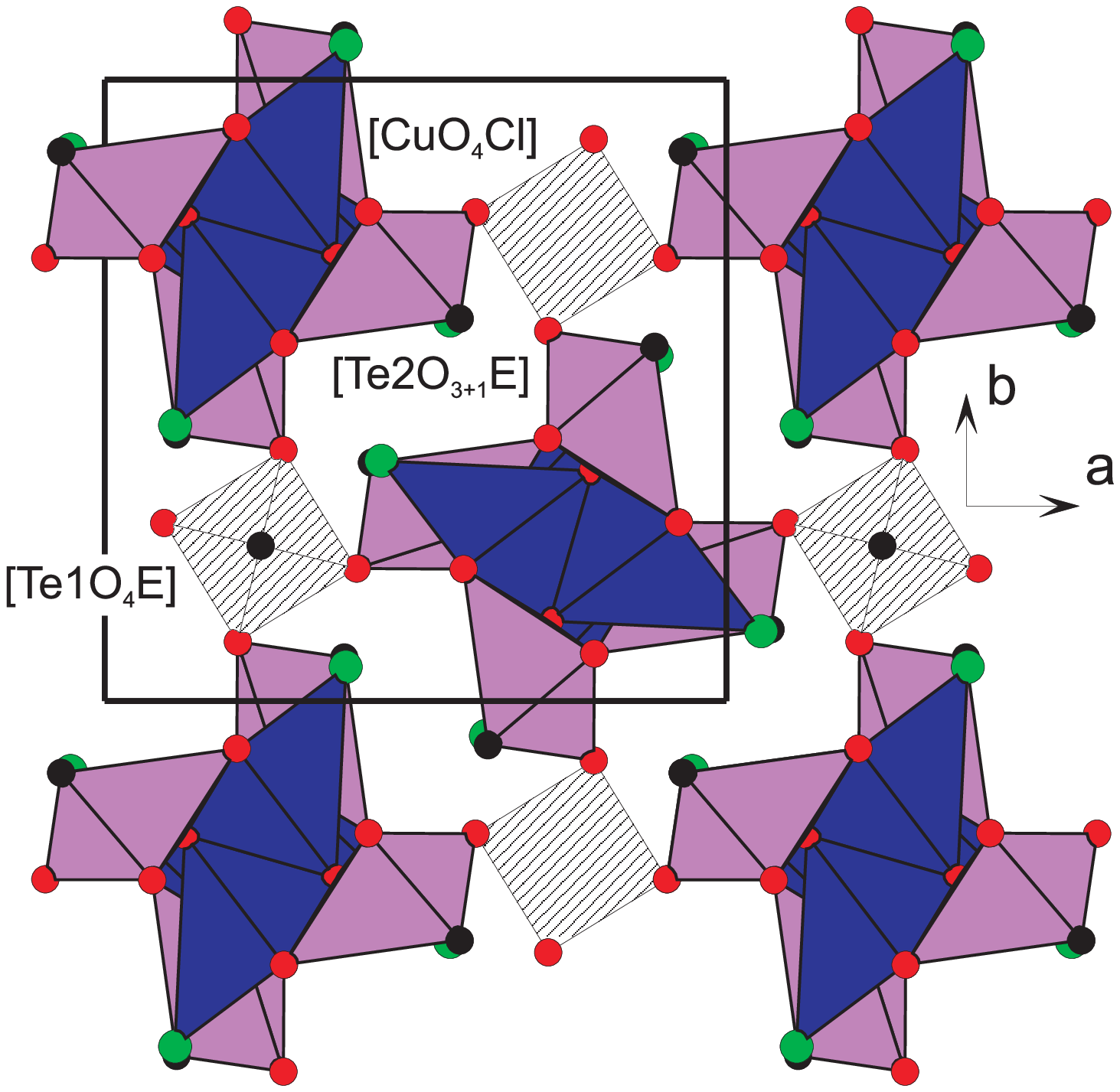}} \caption{(Color online) An overview
of the crystal structure along [001]. } \label{fig3}
\end{figure}

Groups of four [CuO$_{4}$Cl] square pyramids are connected via edge sharing to form
[Cu$_{4}$O$_{8}$Cl$_{4}$] units. These units are arranged such that they form
tetrahedral clusters of Cu$^{2+}$ ions (Figure~\ref{fig2}). The
[Cu$_{4}$O$_{8}$Cl$_{4}$] groups are separated from each other by
[Te(2)$_{4}$O$_{12}$E$_{4}$] rings in the [001] direction and by [Te(1)O$_{4}$E] square
pyramids in the [100] and [010] directions. An overview of the crystal structure is
given in Figure \ref{fig3}.

\section{Comparison with the crystal structure of $Cu-2252$ }
The compounds Cu-2252 and Cu-45124 both crystallize in the tetragonal system but differ
in space groups; \emph{P-4} and \emph{P4/n}, respectively. This implies that Cu-45124 is
centrosymmetric. Nevertheless, the crystal structures of the two compounds have many
similarities. They both contain clusters of Cu(II) ions as [CuO$_{4}$Cl] square pyramids
with similar bond distances and bond angles are connected to form
[Cu$_{4}$O$_{8}$Cl$_{4}$] units. The [TeO$_{3+1}$] coordination around the Te atom in
Cu-2252 is very similar to the coordination around Te(2) in Cu-45124.

\begin{figure}[htbp]
\centerline{\includegraphics[height=5.5cm]{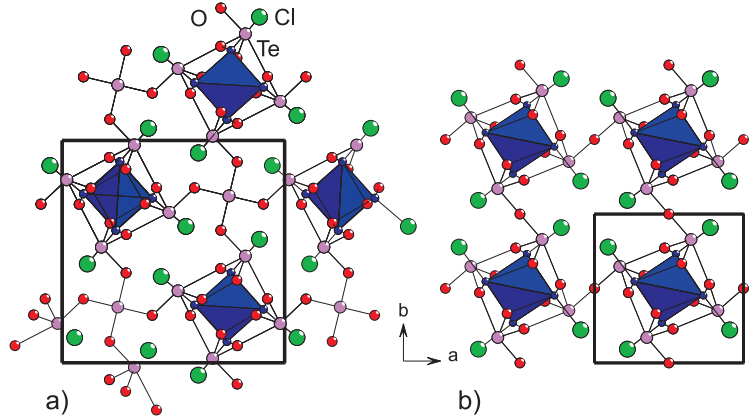}} \caption{(Color online) Crystal
structure  of a) Cu-45124 and b) Cu-2252 projected along [001]. The unit cell is
outlined by the thin line. The figures scale to each other. The structures differ by the
existence of an additional [Te(1)O$_4$] group in the center of the tetrahedra plotted
for Cu-45124. } \label{fig4}
\end{figure}

\begin{figure}[htbp]
\centerline{\includegraphics[height=4.5cm]{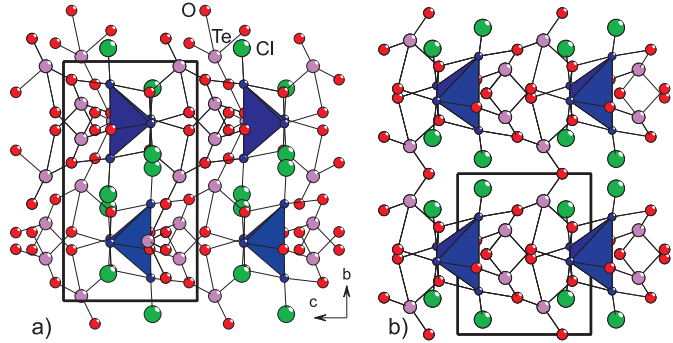}} \caption{(Color online) Crystal
structure in \emph{bc} plane projection of a) Cu-45124 and b) Cu-2252. The figures scale
to each other. } \label{fig5}
\end{figure}

The main difference to Cu-2252 is thus the presence of [Te(1)O$_{4}$] units in Cu-45124.
These units increase the separation of the tetrahedra within the \emph{ab} plane.
Furthermore, they compress a 4-chlorine coordination to a ring which connects the
tetrahedra within the \emph{ab} plane. Thereby, Cl$_4$ and [Te(1)O$_{4}$] units
alternate in Cu-45124 along the \emph{c} axis. In Cu-2252 the corresponding space is
filled up only by tetrahedral Cl$_4$. The \emph{ab} plane projection of the crystal
structure given in Figure \ref{fig4} shows this arrangement in detail. Connected with
this difference of stacking is a different orientation of the Cu-tetrahedra along [001].
For Cu-2252 there is one orientation while there are two different rows of Cu-tetrahedra
in Cu-45124 (see Figure \ref{fig5}). This leads to the space groups being different for
the two compounds.


A band structure study on Cu-2252 with a consequent downfolding and tight binding
analysis has identified the four chlorine sites as the center controlling the in-plane
inter-tetrahedra coupling. The respective hopping matrix element t$_d$ is even
comparable with the intra-tetrahedra hopping t$_1$ \cite{valenti03}. We expect for
Cu-45124 an elongation of the corresponding hopping paths. The in-plane magnetic
exchange should then be reduced compared to the out-of-plane paths and the
intra-tetrahedra exchange. This is also evident from comparing the inter- and intra
tetrahedra Cu -- Cu distances given in Table \ref{cu-distances}. The shortest
inter-tetrahedra distance is elongated by 22\% with respect to Cu-2252, X=Cl.

\begin{table}[htbp]
\caption{Intra- and shortest inter-tetrahedra Cu -- Cu distances.}
\begin{center}
\begin{tabular}{|l|l|l|l|}
\hline
 Compound&
\textbf{Cu}$_{4}$\textbf{Te}$_{5}$\textbf{O}$_{12}$\textbf{Cl}$_{4}$ &
\textbf{Cu}$_{2}$\textbf{Te}$_{2}$\textbf{O}$_{5}$\textbf{Cl}$_{2}$ &
\textbf{Cu}$_{2}$\textbf{Te}$_{2}$\textbf{O}$_{5}$\textbf{Br}$_{2}$  \\
\hline intra-tetrahedra & 3.147 {\AA}& 3.229 {\AA}& 3.196 {\AA}\\
 & 3.523 {\AA} & 3.591 {\AA} & 3.543 {\AA} \\\hline
inter-tetrahedra & 5.063 {\AA}& 4.164 {\AA}&
4.39 {\AA} \\
\hline
\end{tabular}
\label{cu-distances}
\end{center}
\end{table}

\section{Magnetic susceptibility and specific heat}
\begin{figure}[t]
\centerline{\includegraphics[height=7.5cm]{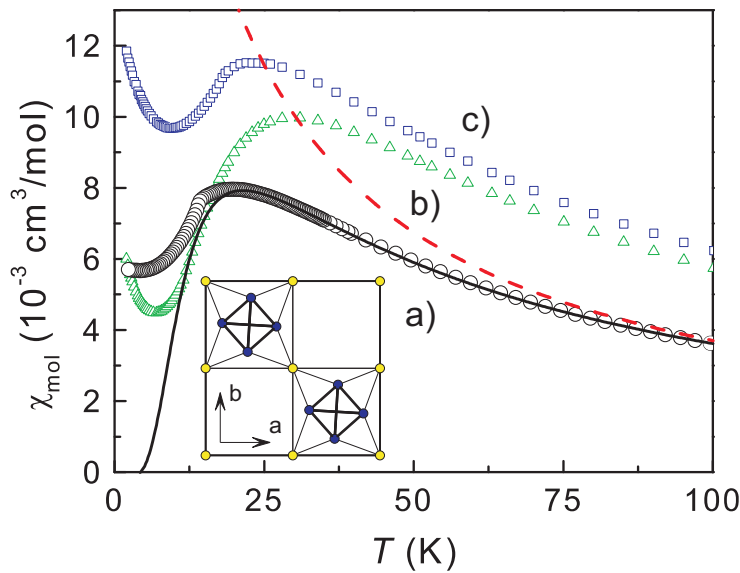}} \caption{(Color online)
Magnetic susceptibility in a field of B=1 Tesla of a) Cu-45124, b) Cu-2251 with X=Br and
c) with X=Cl, respectively \cite{lemmens01}. The solid (dashed) curves give a fit using
the isolated tetrahedron model (the high temperature Curie-Weiss behavior) described in
the text. The insert shows a projection on the Cu tetrahedra (full circles) within a
unit cell and the inversion centers (open circles) in Cu-45124.} \label{sus}
\end{figure}

\begin{figure}[t]
\centerline{\includegraphics[height=9.5cm]{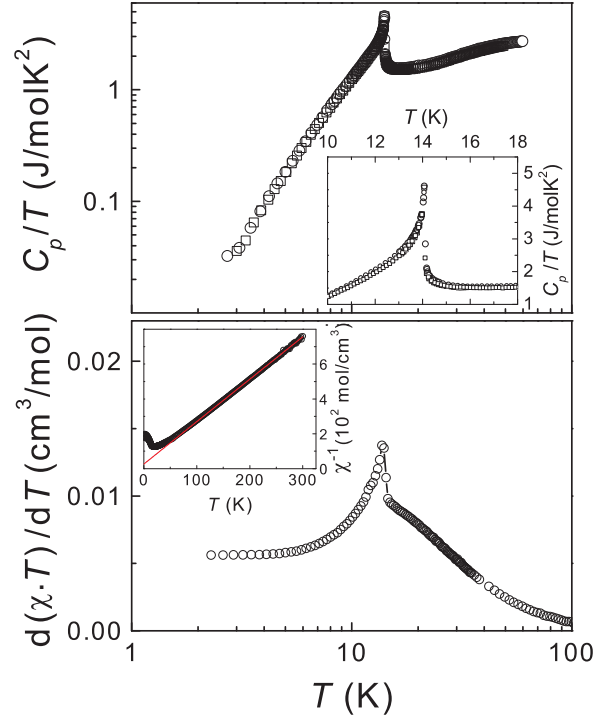}} \caption{(upper panel)Specific
heat $C_p$/T of Cu-45124. The inset zooms into the transition regime. (lower panel)
$d/dT(\chi_{mol}\cdot T$) as function of temperature. The inset shows the reciprocal
magnetic susceptibility.} \label{heat}
\end{figure}

At high temperatures the magnetic susceptibility of Cu-45124 shows a Curie-Weiss
behavior that extrapolates to a paramagnetic Curie-Weiss temperature
$\theta$$_{CW}\approx $-11(1)~K (dashed curve in Fig. \ref{sus} and the lower inset of
Fig. \ref{heat}). The slope of the reciprocal susceptibility corresponds to an effective
magnetic moment of 1.81(1) $\mu_{Bohr}$ per Cu$^{2+}$ ion. With the magnetic moment
arising from Cu$^{2+}$ with $S$=1/2 the effective moment indicates an gyromagnetic ratio
$g \approx$2.1, in fair agreement with the expectation for a 3$d^{\rm 9}$ configuration.
The paramagnetic Curie-Weiss temperature $\theta$ of Cu-45124 is smaller compared to
Cu-2252, with X=Cl ($\theta$$_{CW}\approx $-25~K). It should be highlighted that this
value corresponds to a weighted sum of all inter- and intra-tetrahedra coupling
constants.

The built-up of short range antiferromagnetic correlations appears in Cu-45124 for
temperatures below about 80~K. This is evident as a deviation from the Curie-Weiss
behavior (dashed curve in Fig.~\ref{sus}). Moreover, the correlations are captured
rather well in a model of independent tetrahedra, with two exchange coupling constants,
J$_1$ and J$_2$ which has been detailed in Ref.~\onlinecite{johnsson00}. This is evident
from the solid line in Fig.~\ref{sus} which represents a fit to $\chi_{mol}$ using
J$_1$=32.9~K and J$_2$=18.4~K. For Cu-2252 fits of similar quality have been established
with J$_1$=J$_2$=38.5~K (43K) for X=Cl (Br) \cite{johnsson00}.

At low temperature the susceptibility shows a broad maximum at $T_{max}$=29~K and a kink
below $\approx$14K identified with long range ordering. There is no indication for a
thermal hysteresis when comparing the field-cooled and zero-field-cooled measurement.
Taking the derivative with respect to temperature\cite{Kremer88} of the product
$\chi_{mol}\cdot T$ the kink shows up as a $\lambda$-type anomaly at T$_c$ similar to
that observed in the heat capacity as shown in Figure \ref{heat}. The quantity
$d/dT(\chi_{mol}\cdot T$) falls off with a long tail indicating substantial short-range
antiferromagnetic correlations toward higher temperatures. The specific heat plotted in
the upper part of Fig.~\ref{heat} can be used to determine the magnetic entropy
$S_{mag}(T)$ after subtracting an estimated lattice contribution. $S$(14K) contains an
essential fraction of the magnetic entropy released in the phase transition and amounts
to about 40 {\%} of R$\cdot$ln2, expected for a S=1/2 system. This value is a little bit
larger than for Cu-2252 with X=Cl.

Characteristic temperatures and energy scales are given in Table \ref{sus-data} and
compared with the data from Cu-2252. Although the overall magnetic behavior is very
similar, Cu-45124 behaves more close to the corresponding chlorine Cu-2252 (X=Cl). We
can use the maximum in the susceptibility to test or establish some scaling. It is
evident from a comparison of T$_c$/T$_{max}$ for different compounds that the
composition and the related changes of the inter-tetrahedral coupling mainly effect the
transition temperature T$_c$. The entropy at the transition determined from specific
heat and the ordered moment from neutron scattering follow roughly the behavior of
T$_c$/T$_{max}$. In contrast, T$_{max}$ itself and the maximum in the magnetic Raman
scattering E$_m$, discussed further below, are dominated by intra-tetrahedra couplings
and show no obvious scaling.

\begin{table}[htbp]
\begin{center}
\caption{Data derived from magnetic susceptibility (T$_c$, T$_{max}$), specific heat
(entropy $S(T_c)$) and Raman scattering (E$_m$, $\Delta$E) compared with data for the
other tetrahedron systems derived from Ref.
\onlinecite{lemmens01,lemmens02-proceeding,jensen03}. The ordered magnetic moment
$\mu_{ord.}$/$\mu_B$ determined by neutron scattering is from Ref.
\onlinecite{zaharko04,zaharko05}. For X=Br the parameter $\Delta$\textit{E} corresponds
to the FWHM of the continuum. Further details see text.}

\begin{tabular}{|c|c|c|c|}
  \hline
  Compound & Cu$_{2}$Te$_{2}$O$_{5}$Cl$_{2}$ & Cu$_{4}$Te$_{5}$O$_{12}$Cl$_{4}$ & Cu$_{2}$Te$_{2}$O$_{5}$Br$_{2}$ \\
  \hline\hline
  $T_{max}$/K               & 23 & 19 & 30 \\
  $T_{c}$/K                 & 18.2 & 13.6 & 11.4 \\
  $T_{c}$/T$_{max}$         & 0.79 & 0.72 & 0.38 \\ \hline
  $S(T_{c})$/(R$\cdot ln2)$ & 0.36 & 0.4 & 0.16 \\
  $\mu_{ord.}$/$\mu_B$      & 0.7-0.9  & - & 0.4 \\ \hline
  E$_{m}$/cm$^{-1}$         & 47.5 & 36.9 & 60 \\
  $\Delta$E/cm$^{-1}$       & 9.0 & 4.7 & 22 \\
  $E_{m}$/$T_{c}$           & 3.8 & 3.9 & 7.6 \\
  $E_m$/$T_{max}$           & 3.0 & 2.8 & 2.9 \\
  \hline
\end{tabular}
\label{sus-data}
\end{center}
\end{table}

The magnetic susceptibility of Cu-45124 differs from the other two Cu-2252 systems in
the sense that there is no upturn at low temperatures. This also corresponds to a
different low-temperature curvature. Such an upturn can have intrinsic or extrinsic
origins. Defects may induce paramagnetic centers that show up in the susceptibility as
an additional low temperature contribution with a Curie-Weiss-like temperature
dependence \cite{laukamp97,fischer98c,manabe98}. As the discussed compounds have very
similar chemical properties and are prepared from the same starting materials via very
similar preparation routes, however, there is no reason to assume a fundamental
different defect density. As a possible reason we suggest the existence of an inversion
center and the higher symmetry of Cu-45124 \cite{symmetry}. The inset in
Figure~\ref{sus} shows a projection on the unit cell including the Cu tetrahedra and
respective inversion centers. Exchange paths that are inversion symmetric do not allow
antisymmetric spin-spin interactions, as the DM interaction. Therefore, staggered fields
\cite{affleck99} and effective ferromagnetic moments at low temperatures are suppressed.
In $\rm Sr_2V_3O_9$ and $\rm Ba_2V_3O_9$ low temperature upturns in the magnetic
susceptibility have been attributed to low symmetry exchange paths \cite{kaul03}. We are
aware that this argument might only be relevant for inter-tetrahedra exchange and is
qualitative as it cannot be proven for all possible exchange paths individually.
However, DM interactions that are allowed for Cu-2252 have been shown to be of relevance
for the stabilization of ground states with small ordered moments
\cite{jensen03,kotov05}.

\section{Raman scattering}
The overall Raman scattering intensity in Cu-45124 is very large, compared to transition
metal oxides \cite{lemmens-rev}. This is based on the enormous electronic polarizability
and nonlinearity of oxo-tellurides that makes them promising materials for
photorefractive, acousto-optical or applications related to second harmonic generation
\cite{domoratskii00}. Due to the inversion center, however, photorefractive effects are
not allowed in Cu-45124. Also certain phonon-phonon and spin-phonon scattering terms are
forbidden. Furthermore, possible nonlinearities do not break symmetry selection rules
and there exist no further evidence for a structural or electronic instability as will
be shown below.

\begin{figure}[t]
\centerline{\includegraphics[height=7.5cm]{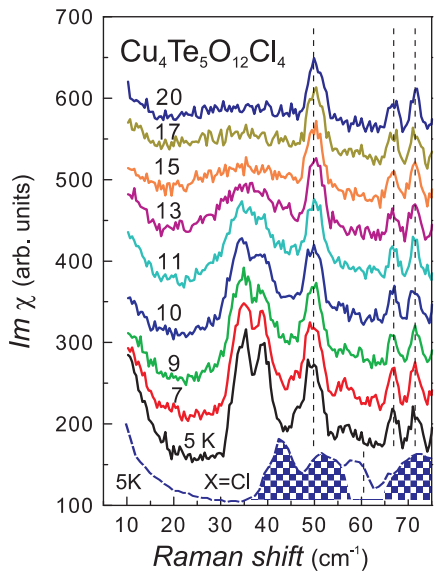}} \caption{(Color online) Bose
corrected Raman scattering intensity of Cu-45124 at small energies and low temperatures.
Phonon modes are marked by dashed lines. The curves have been given a vertical offset
for clarity. The curve at the bottom corresponds to Cu-2252 with X=Cl. The three shaded
maxima are the magnetic Raman intensity. A phonon with an asymmetric Fano lineshape is
omitted. } \label{raman-low}
\end{figure}

We observe 46 sharp modes in the frequency regime 40 -- 800~cm$^{-1}$ that do not show a
strong or anomalous temperature dependence. Therefore they are attributed to optical
phonon modes. In contrast, the low energy regime exhibits modes with frequencies of 30
-- 65~cm$^{-1}$ that evolve in the low temperature regime where the susceptibility
changes, see Figure \ref{raman-low}.

\begin{figure}[t]
\centerline{\includegraphics[height=6.5cm]{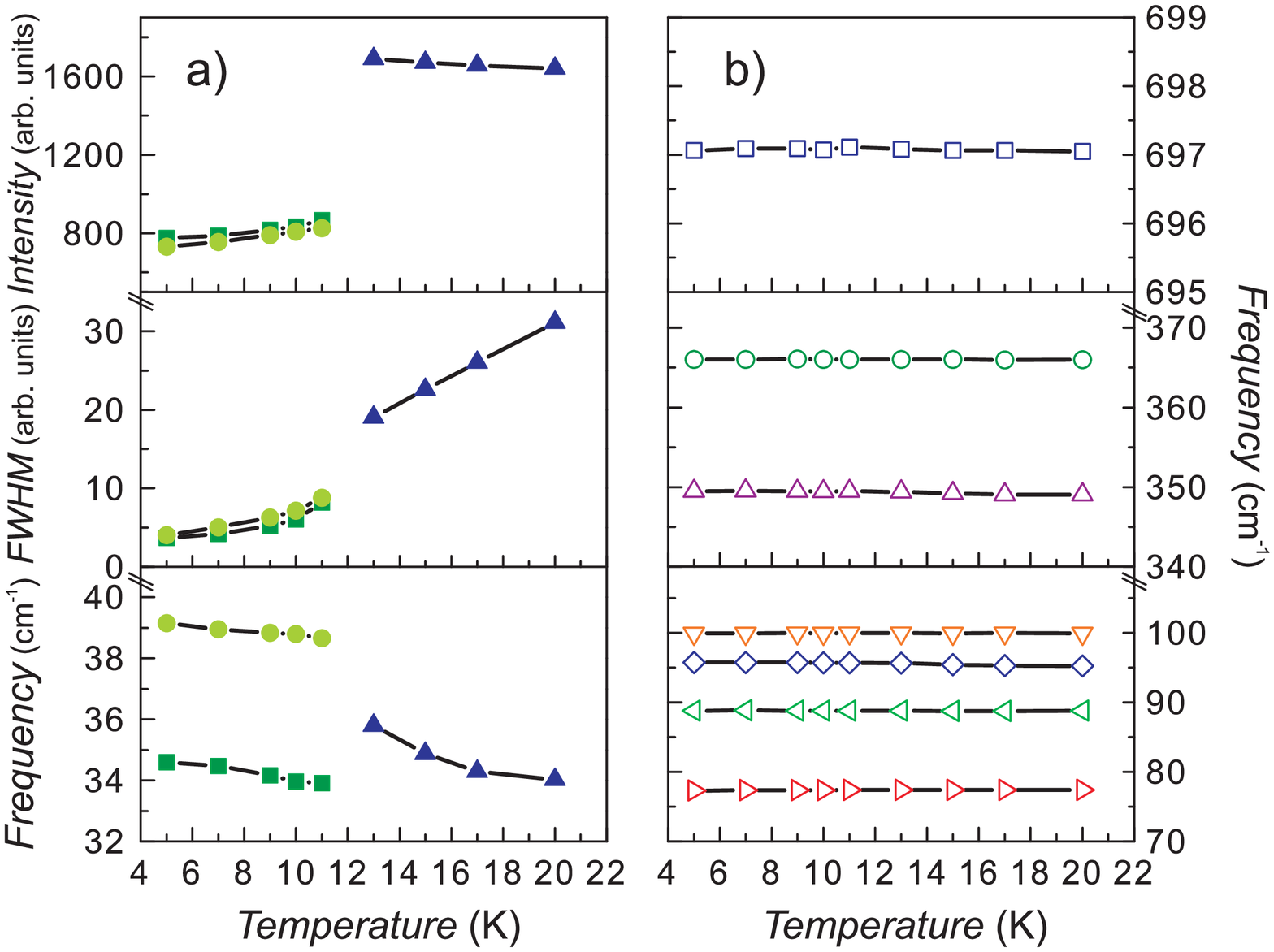}} \caption{(Color online)
Analysis of the temperature dependence of the a) low energy modes and b) phonon modes in
Raman scattering. } \label{raman-analysis}
\end{figure}

Based on the crystallographic coordinates given in Table \ref{coordinates} a symmetry
analysis leads to $\Gamma^{8g}$=(3Ag +3Au+3Bg+3Bu+3Eg+3Eu) modes for each of the 6
\emph{8g} sites and $\Gamma^{2c}$=Ag+Au+Eg+Eu modes for the \emph{2c} site. In total
$\Gamma_{even}$=58 modes are Raman active. This number is in reasonable agreement with
the observed 46 modes keeping in mind that due to a near degeneracy or small intensity a
few modes can be covered. In Figure \ref{raman-analysis}b) the result of a temperature
analysis of representative phonons is shown. The transition does not show up in the
phonons, i.e. there is no evidence for pronounced spin-phonon coupling. The same
observation has been made for Cu-2215 using Raman and IR spectroscopy
\cite{lemmens01,perucchi04}. It disagrees with conclusions drawn from thermal
conductivity measurements on the latter compound \cite{prester04}.

At small Raman shifts there are two modes (34.5~cm$^{-1}$, 39.2~cm$^{-1}$) that show a
pronounced temperature dependence of the intensity close to T$_c$ and one weaker maximum
(58~cm$^{-1}$), see Figure \ref{raman-low}. The analysis of this data is given in Figure
\ref{raman-analysis}a). The main modes develop from a very broad maximum that gradually
shifts to higher energy and then splits-up equidistantly right at the transition. In the
same temperature regime the linewidth of the modes strongly decreases. We do not observe
quasielastic scattering at elevated temperatures that is common for low dimensional spin
systems \cite{lemmens-rev}. In contrast, a tiny increase of intensity is seen at lowest
temperatures. We attribute these modes to magnetic scattering as their energy and
temperature dependence match the related energy scales. The bottom curve in Figure
\ref{raman-low} shows the corresponding data for Cu-2252 with X=Cl. There are two
similar maxima at 43 and 52~cm$^{-1}$, a phonon at $\approx$60~cm$^{-1}$ and a third
magnetic mode at 73~cm$^{-1}$.


To characterize the magnetic scattering on a molecular field like level we determine the
mean frequency E$_{m}$=36.9~cm$^{-1}$ and the splitting of the modes
$\Delta$E=4.7~cm$^{-1}$ for Cu-45124. It is obvious from Figure \ref{raman-analysis}
that E$_{m}$, characteristic for a magnetic energy scale, is only weakly temperature
dependent. In contrast, $\Delta$E, increases rapidly at T=T$_{c}$ and is constant to
lower temperatures. The temperature dependence is even sharper than expected for a
magnetic order parameter.

\section{Discussion}
The ambivalence of weakly coupled spin cluster systems in the proximity to a quantum
critical point is an interesting topic and it has been shown that both thermodynamic and
spectroscopic tools have to be used to understand their properties
\cite{kotov99,kotov00,brenig01,kotov05}. The presently studied tetrahedra systems fall
into the rare case where both local and collective excitations can be observed
simultaneously, and where the character of the excitation spectrum changes with minute
changes of external parameters.

In the limit of weak coupling, i.e. in the quantum disordered phase, a set of gapped low
energy excitations exists with energies characteristic for the intra-tetrahedra or
-dimer couplings. Such modes have been observed using different spectroscopic tools,
e.g. in the frustrated dimer system $\rm SrCu_2(BO_3)_2$
\cite{kageyama99,lemmens00,nojiri03,aso05,gozar05}. In the limit of strong
inter-tetrahedra and weak intra-tetrahedra coupling long range ordering leads to gapless
collective modes. In Raman scattering such modes are observed as broadened two-particle
continua with energies up to a few times the coupling constant. Such a scattering
contribution survives elevated temperatures with respect to its integrated intensity.
However, it softens considerably to lower energies and forms quasi-elastic tails.
Theoretical modeling of two-particle Raman scattering on spin tetrahedra systems showed
symmetric continua only if the inter-tetrahedra couplings do not dominate, e.g. due to a
coupling to chains of tetrahedra \cite{brenig01,brenig02-proceeding,brenig03}. In such
models the determination of a mean energy of the continuum E$_{m}$ is meaningful as its
position is mainly determined by the strong intra-tetrahedra couplings.

This allows us to derive a correspondence between the Raman data and other parameters of
the systems. The bottom of Table \ref{sus-data} shows a synopsis for all tetrahedra
systems\cite{jensen03,lemmens02b,choi06}. In Cu-2252 with X=Br the broad, symmetrical
continuum leads to a very large E$_{m}$=60~cm$^{-1}$. From a comparison with the other
systems it is clear that this parameter scales with the maximum position in the
susceptibility. We attribute changes of E$_{m}$ to modulations of the intra-tetrahedra
coupling. Only for Cu-2252 with X=Br, the system with the smallest inter-tetrahedra
coupling, the strongly reduced ordered moment shows longitudinal fluctuations
\cite{tun90,zheludev02} observed as a distinct longitudinal magnon mode at 18~cm$^{-1}$
(Refs. \onlinecite{gros03} and \onlinecite{jensen03}). The energy separation of the
continuum to this mode is of the order of the intra tetrahedra coupling
\cite{lemmens01,gros03} and should be an indication for the proximity to a quantum
critical point.

In contrast, Cu-2252 with X=Cl and even more Cu-45124 show splittings with considerably
smaller energies $\Delta$E. We propose two-spin anisotropies to be responsible for these
effects as the weaker inter-tetrahedra coupling should lead to broader signals and
single ion anisotropies do not exist for $\rm Cu^{2+}$ with s=1/2. Including DM
interaction into the Hamiltonian of SrCu$_2$(BO$_3$)$_2$ a very satisfactory description
of Raman modes and splittings of ESR lines have been accomplished \cite{gozar05}. For
Cu-2252 with X=Br the field dependence of the longitudinal magnon and T$_{c}$ has been
modeled \cite{jensen03}. Within this approach we would expect a smaller contribution of
the DM interaction as the normalized splitting is $\Delta$E/E$_{m}$=12.7 for Cu-45124
compared to $\Delta$E/E$_{m}$=19 for Cu-2252 with X=Cl. This is consistent with the
behavior of the low temperature susceptibility.

The structural difference between the compounds discussed in Chapter IV are summarized
as a reduction of the inplane inter-tetrahedra coupling introducing additional
[Te(1)O$_{4}$] groups. The different stacking of the tetrahedra and the inversion center
are expected to affect the out-of-plane exchange, however, only to a minor degree. The
question remains which scenario dominates the spin fluctuations in Cu-45124 irrespective
of the exact ground state. Is the dimensionality enhanced by a decreasing in-plane
exchange with respect to out-of-plane exchange? As a result the character of the
transition should be more mean field-like. Or does the same reduction of the in-plane
exchange enhances the effect of frustration of the intra tetrahedra exchange? Although
in Cu-45124 the small number of low energy excitations might be taken as evidence for
competing interactions, the reduced transition temperature while keeping the entropy
constant proposes that the change of dimensionality is more effective for Cu-45124.
Noticeably, recent {\it ab initio} calculations on this new tellurate compound support
the mean field nature of the magnetic behavior as observed experimentally
\cite{roser06}.

Neutron scattering experiments on Cu-2252 show that a complex helical state with a
reduced ordered moment can be used to describe the ordered state
\cite{zaharko04,zaharko05}. Inelastic scattering detects two kind of excitations, a
dispersionless high energy mode at $\approx$48~cm$^{-1}$ and a dispersing mode with a
gap of $\approx$16~cm$^{-1}$. Remarkably, for X=Cl the dispersionless mode survives to
elevated temperatures and is only moderately depressed for T$>$T$_c$ \cite{streule06}.
These observations support our results and we expect a similar trend for Cu-45124.
Further Raman scattering and thermodynamic experiments are prepared to test whether
controlled substitutions or pressure can be used to shift Cu-45124 more closer to the
quantum critical point similar to observations in Cu-2252 \cite{lemmens02b,kreitlow05}.


\section{Conclusions}
The compound Cu-45124 has been established as the second example of a system with weakly
interacting s=1/2 spin tetrahedra. Results from Raman scattering and thermodynamic
experiments have been compared with the related compounds and discussed in terms of
simple scaling arguments. We conclude a reduced effect of competing interactions and a
mean field like behavior in the present system compared to the intensively investigated
system Cu-2252.

\section{Acknowledgements}
This work has been carried out with financial support from the Swedish Research Council,
the German Science Foundation and the ESF program \emph{Highly Frustrated Magnetism}. We
acknowledge important discussions with R. Valent$\acute{i}$.




\end{document}